\documentclass[11pt]{article} 
\date{\today}

\usepackage[utf8]{inputenc}
\usepackage{graphicx}
\usepackage{amsmath}
\usepackage{breqn}
\usepackage{afterpage}
\usepackage{blindtext}
\usepackage{amssymb}
\usepackage{float}
\usepackage{hyperref}
\usepackage{authblk}
\usepackage{geometry}
\usepackage{authblk}

\title{Search for dark matter towards the Galactic Centre with 11 years of ANTARES data}
\author[1,2]{A.~Albert}
\author[3]{M.~Andr\'e}
\author[4]{M.~Anghinolfi}
\author[5]{G.~Anton}
\author[6]{M.~Ardid}
\author[7]{J.-J.~Aubert}
\author[8]{J.~Aublin}
\author[8]{B.~Baret}
\author[9]{S.~Basa}
\author[10]{B.~Belhorma}
% ...

\affil[1]{\scriptsize{Universit\'e de Strasbourg, CNRS,  IPHC UMR 7178, F-67000 Strasbourg, France}}
\affil[2]{\scriptsize{Universit\'e de Haute Alsace, F-68200 Mulhouse, France}}
\affil[3]{\scriptsize{Technical University of Catalonia, Laboratory of Applied Bioacoustics, Rambla Exposici\'o, 08800 Vilanova i la Geltr\'u, Barcelona, Spain}}
\affil[4]{\scriptsize{INFN - Sezione di Genova, Via Dodecaneso 33, 16146 Genova, Italy}}
\affil[5]{\scriptsize{Friedrich-Alexander-Universit\"at Erlangen-N\"urnberg, Erlangen Centre for Astroparticle Physics, Erwin-Rommel-Str. 1, 91058 Erlangen, Germany}}
\affil[6]{\scriptsize{Institut d'Investigaci\'o per a la Gesti\'o Integrada de les Zones Costaneres (IGIC) - Universitat Polit\`ecnica de Val\`encia. C/  Paranimf 1, 46730 Gandia, Spain}}
\affil[7]{\scriptsize{Aix Marseille Univ, CNRS/IN2P3, CPPM, Marseille, France}}
\affil[8]{\scriptsize{Universit\'e de Paris, CNRS, Astroparticule et Cosmologie, F-75013 Paris, France}}
\affil[9]{\scriptsize{Aix Marseille Univ, CNRS, CNES, LAM, Marseille, France }}
\affil[10]{\scriptsize{National Center for Energy Sciences and Nuclear Techniques, B.P.1382, R. P.10001 Rabat, Morocco}}
% ...

\begin{document}

\maketitle 
\newpage
\begin{abstract}

Neutrino detectors participate in the indirect search for the fundamental constituents of dark matter (DM) in form of weakly interacting massive particles (WIMPs).
In WIMP scenarios, candidate DM particles can pair-annihilate into Standard Model products, yielding considerable fluxes of high-energy neutrinos. A detector like ANTARES, located in the Northern Hemisphere, 
is able to perform a complementary search looking towards the Galactic Centre, where a high density of dark matter is thought to accumulate. Both this directional information and the spectral features of annihilating DM pairs are entered into an unbinned likelihood method to scan the data set in search for DM-like signals in ANTARES data. 
Results obtained upon unblinding 3170 days of data reconstructed with updated methods are presented, which provides a larger, and more accurate, data set than a previously published result using 2101 days.
A non-observation of dark matter is converted into limits on the velocity-averaged cross section for WIMP pair annihilation.  
\end{abstract}


\section{Introduction: dark matter signals at neutrino telescopes}

The existence of cold, non-baryonic dark matter (DM), evidenced on macroscopic scale by astrophysical observations  \cite{PDG}, encourages the searches for its possible particle constituents. 
Among those candidates, most WIMP scenarios accommodate the DM relic density reported by astrophysical measurements through a freeze-out mechanism.
This could imply that typical WIMP interactions of the DM candidate, especially its annihilation cross section,  lie near the electroweak scale; beyond that, other parameters 
like the candidate WIMP mass or the specific details of the DM model are left unbound. 
Under the hypothesis that a WIMP coincides with its antiparticle, indirect searches for WIMPs are possible by 
detecting a signature of WIMP annihilation into Standard Model particles.
Such signals are therefore searched from the direction of massive astrophysical environments, where	WIMPs can be gravitationally attracted.
DM builds up in and around massive celestial bodies and gravitational accumulators, and is organized in {\em{halos}} and {\em{clumps}}. The distribution of dark matter with density $\rho$ at a given sky location ($r, \theta, \phi$) is described through the $J$-factor 
\begin{equation}
    \label{eqJ}
    J = \int_\Omega d\Omega(\theta,\phi) \int_{\rm l.o.s.} \rho^2\left(s(r,\theta,\phi)\right) ds,
\end{equation}
with $\Omega$ being the solid angle under which the source is observed, and $s$ the radial coordinate integrated over the line of sight (${\rm l.o.s.}$) (see \cite{JF} for a detailed discussion). For neutrino telescopes, which have a very broad field of view, values as large as $10^\circ - 30^\circ$ can be considered for the opening angle characterising the solid angle $\Omega$. 
Preferred locations where dark matter is predicted to accumulate are:
\begin{enumerate}
    \item the Galactic Centre, having the largest $J$-factor;
    \item massive, non-luminous galaxies like dwarf spheroidals;
    \item the Sun or other nearby very massive celestial bodies.
\end{enumerate}
DM messengers for indirect searches are neutrinos, $\gamma$ rays or charged cosmic rays ($e^+$, $\bar{p}$), produced either as primary or as secondary products of a WIMP pair annihilation, through different channels.
The Galactic Centre is not only a promising source for its large predicted DM density; it is also a target of complementary searches for neutrino detectors and $\gamma$-ray telescopes, due to the low source contamination that would give way to an unambiguous signal identification.
Lastly, the Galactic Centre is in good visibility for neutrino telescopes located in the Northern Hemisphere (as will be clarified in Section \ref{sec2}), or for $\gamma$-ray telescopes installed in the Southern Hemisphere. 
The flux of 
neutrinos reaching the Earth from a WIMP pair annihilation can be expressed as
a function of the thermally averaged cross section $\langle \sigma v \rangle$ for WIMP pair annihilation,
of the energy distribution of outcoming particles per WIMP pair collision $dN/dE_\nu$, and of the DM distribution represented by the $J$-factor:
\begin{equation}
\label{flux}
    \frac{d\Phi(E_\nu)}{dE_\nu} = \frac{1}{4 \pi M_{\mbox{\tiny WIMP}}^2}\frac{\langle \sigma v \rangle}{2} \frac{dN(E_\nu)}{dE_\nu} J,
\end{equation}
where the factor 1/2, used in this analysis, holds for self-conjugate WIMPs, and is to be replaced by a factor 1/4 otherwise.
Similarly, the term $1/M_{\mbox{\tiny WIMP}}^2$ 
arises from the presence of two WIMPS in the process, keeping into account that both the mass and the volumetric
density are expressed in energy units.
Through the relation in Equation~(\ref{flux}), a measurement of the integrated neutrino and antineutrino flux from the region of the Galactic Centre
\begin{equation}
\Phi_{\nu+{\bar{\nu}}} = \int dE_\nu \frac{d\Phi_{\nu}}{dE_\nu}+ \int dE_{\bar{\nu}} \frac{d\Phi_{\bar{\nu}}}{{dE_{\bar{\nu}}}}
\end{equation}
is converted into limits on the thermally averaged cross section $\langle \sigma v \rangle$ for WIMP pair annihilation. 
A lower bound on this quantity of $3\cdot10^{-26}$ cm$^3$ s$^{-1}$ can be made based upon cosmology arguments \cite{Bertone}.

\subsection{Directional and morphological information}
\label{sec11}

Indirect searches for dark matter are unavoidably subject to large uncertainties, mostly arising from the parameterisation of the unknown DM distribution. The spherically averaged DM density profile $\rho$ contained in the $J$-factor (Equation~(\ref{eqJ})) is modelled according to different assumptions, leading to considerably different results.
The main assumptions on $\rho$ are based on cosmological N-body simulation results and/or dynamical constraints on the Milky Way or spiral galaxies.
Even if baryonic physics (star formation and feedbacks) is not fully under control in hydrodynamics simulations, the baryons may steepen or even flatten the inner behaviour of the DM profile (see e.g. \cite{2016MNRAS.456.3542T,2014Natur.506..171P,Mollitor:2014ara,1986ApJ...301...27B}).
Alternatively, dynamical studies of galaxies show a large diversity in rotation curves ~\cite{2019A&A...623A.123G} and can suggest a cored DM profile~\cite{2010AdAst2010E...5D,Famaey:2015bba}.
A popular and simple parameterisation of the DM density obtained in pure (without baryons) DM cosmological simulations is the Navarro-Frenk-White (NFW) profile \cite{NFW}:

\begin{equation}
     \rho_{NFW}(r) = \frac{\rho_0}{\frac{r}{r_s}\left( 1+ \frac{r}{r_s}\right)^\gamma}
    \label{eq:JNFW}
\end{equation}
with  $\gamma=2$.  
The NFW profile is adopted in the present analysis with $\rho_0 = 1.40\cdot 10^{7}\,\mbox{M}_{\odot}/\mbox{kpc}^3$ and $r_s = 16.1 \mbox{ kpc}$ \cite{NestiSallucci}. 
For the sake of illustrating those DM density uncertainties, other two cases are considered: the profile from the recent study of McMillan \cite{McMillan} giving an internal power law $r^{0.79\pm 0.32}$, and the Burkert profile \cite{Burkert} for which the inner density is constant.

\subsection{Energy Information}
The energy distribution of a neutral massive particle pair-annihilating into Standard Model products can be effectively described with a Monte Carlo generator such as PYTHIA or HERWIG \cite{PYTHIA,HERWIG}. 
The PPPC4 {\em{cookbook}} \cite{PPPC4}, used in this analysis, directly provides spectra for WIMP annihilations into Standard Model modes which are straightforward to adapt to any kind of indirect searches. \\
PPPC4 yields the energy distribution for an isotropic flux of Standard Model particles
originated	in	the	
WIMP	pair	annihilation	at	the	source.
Several final states of the annihilation process, resulting in different decay modes ($\tau^+ \tau^-$, $W^+W^-$, $b\bar{b}$, $\mu^+\mu^-$, $\nu\bar{\nu}$) have been simulated, evaluating
the spectrum of the resulting neutrino flux, $dN_\nu/dE_\nu$,
for each WIMP mass. Each channel is considered with a 100\% branching ratio (BR).
Note that the matter density in the Galactic Centre is not enough to cause distortions or absorption effects in outcoming neutrino spectra.

Flavour oscillations occur between source and detection point.
The three neutrino flavours are equally produced in WIMP pair annihilations, and the data set considered here only contains muon neutrinos recorded at the detector. The oscillations $\nu_e, \nu_\tau$ into $\nu_\mu$, as well as
the loss of $\nu_\mu$ into the other two flavours, have been accounted for.
The energy distribution of neutrino final states is therefore obtained from a modulated superposition of the three flavours, in the long-baseline 
approximation\footnote{The $E/L$ dependency of the oscillations are averaged out for GeV--TeV neutrino energies over the distance between the Earth and the Galactic Centre.}, with coefficients taken from \cite{PDG}.
Neutrinos and antineutrinos are symmetrically produced in  WIMP annihilations, and are detected indistinctly
by current neutrino telescopes. This analysis is restricted to muon neutrino events at the detector, as will be described in Section \ref{sec2}.

\section{Detector and Data Set}
\label{sec2}
ANTARES is an underwater Cherenkov detector situated in the Mediterranean Sea 40 km offshore from Toulon. It is composed of 12 detection lines instrumented with photomultiplier tubes enclosed in optical modules \cite{ANTARES}. 
ANTARES	data	analysis	allows for	energy	and	directional reconstruction	of	charged particle tracks	originated	from	a	neutrino	
interaction	occurring	around	the	detector.
The very large background of muons produced in atmospheric interactions of cosmic rays is suppressed by considering events with arrival directions crossing the Earth.
Under this condition, the Galactic Centre, located at a declination of $-29.01^\circ$, is visible from the detector latitude about 70$\%$ of the time \cite{Visibility}.

In this paper, 11 years of data collected with ANTARES between May 2007 and December 2017 are analysed, updating upon prior searches 
\cite{ANTARES_GCWIMP}.
Signatures of neutrinos from DM annihilation are searched for in a data sample composed of reconstructed muon tracks originating from charged current (CC) interactions of neutrinos around the detector. 
A set of pre-selection cuts has been applied to discriminate these $\nu_\mu$ CC-induced events from atmospheric muon background; this first discrimination is based on the zenith angle of provenience of the event and on the quality of the track reconstruction. 
Tracks are reconstructed in ANTARES from the position and times of photomultiplier hits, recorded in general from different detector lines.
The quality parameter is, in the standard approach, a maximum likelihood $\Lambda$ obtained with a multi-line reconstruction fit \cite{AAFit}. 
At low energies, however, it is possible to best reconstruct those tracks hitting only one line of the detector using a single-line reconstruction \cite{BBFit}; 
this fit is based on a $\chi^2$ minimization and the $\chi^2$ value serves as a quality parameter. The single-line reconstruction is more efficient for energies below $\sim$~100 GeV.

The parameters $\Lambda$ and $\chi^2$ are used as quality indicator for multi-line and single-line tracks respectively. Additionally, an angular error estimate $\beta$, provided by the multi-line reconstruction fit, 
has been considered. Variable cuts have been applied as reported in Table \ref{tablecut}, and the values yielding best sensitivity have been chosen to unblind the data, as explained later in section \ref{sectionsens}.
\begin{table}[h!]
\centering

\begin{tabular}{|ll|}
\hline
 Fit&  Cut value\\
 \hline
 Multi-line & $\Lambda > -5.2 $ \\
  Multi-line & $\beta < $ 1$^\circ$ \\
 Single-line & $\chi^2 <  0.7$\\
Both & $\cos\theta > 0 $  \\
 \hline 
  \end{tabular}

\caption{
Final selection criteria applied to the data set. The quality of the multi-line reconstruction fit is evaluated by a likelihood $\Lambda$ and angular error estimate $\beta$; 
$\chi^2$ characterises the single-line fit; the angle $\theta$ 
is complementary to the zenith, such that $\cos\theta>0$ identifies an {\em upgoing} track, coming from across the Earth.}
\label{tablecut}
\end{table}

This sample is composed of 8976 tracks reconstructed with the multi-line algorithm and 2522 tracks with the single-line algorithm recorded over 3170 days of effective livetime; note that in the text that follows the term neutrinos stands for $\nu+{\bar{\nu}}$, 
as the	events	generated	by	their	
interactions	are	seen	indistinguishably in current neutrino telescopes.
Tracks are reconstructed with an angular resolution of the order of 1$^\circ$ at the energies relevant for this search \cite{AntaresAF}.
Given	its	geometry	and	volume,	the	ANTARES	telescope	is	optimised for the detection of	neutrinos	with	energies	from	about	20	GeV	to a few	PeV.		The	DM analysis	is,	therefore,	
in	the	medium WIMP	mass range.
The	amount of Cherenkov	photons	induced	along the paths of the propagating charged particles is	proportional	to	the	amount of deposited energy	and, consequently, the	number	of	hit	optical	modules,  $N_{\mbox{\tiny{HITS}}}$,	 is	
a	good	proxy	of	the neutrino energy	$E_\nu$ .

A set of simulated data has been produced in correspondence with the environmental and trigger conditions of each data run~\cite{AntaresSIM}, and has been adapted to the specific DM analysis through the use of weights reproducing the energy distribution $dN_\nu/dE_\nu$ of each WIMP annihilation channel. 
The simulated data used for this search contain $\nu_\mu$ CC induced muons; the contribution of muons from $\nu_\tau\rightarrow\tau$ and subsequent $\tau$ decay is not considered in the simulated sample used in this analysis.

The search is optimised on shuffled ($blind$) right-ascension data, which are unblinded after having established the best selection criteria. 
A newly released version of the ANTARES reconstruction software \cite{AAFit,BBFit} was run on the full data set.
With the new processing and reconstruction of the data a considerable amount of livetime could be recovered with respect to the  previous 9-year study \cite{ANTARES_GCWIMP}.

The search method used for this analysis is the same as that used in the previous study \cite{ANTARES_GCWIMP}, keeping into account the correction of a computation problem which affected 
the previous results~\cite{ANTARES_GCWIMP}.


\section{Method}
The signal from DM annihilation is expected to appear as a cluster of neutrino events scattered around the position of the Galactic Centre according to the $J$-factor profile, whose energy distribution reproduces the WIMP annihilation spectra \cite{PPPC4}. 
This spatial cluster of signal events is to be found over a background of atmospheric neutrinos~\cite{Honda}.
Both background estimation and search optimisation use shuffled (blinded) real data, by replacing the right
ascension value with a random value between
0$^\circ$ and 360$^\circ$. 
This random shuffle washes out any possible spatial clustering in correspondence to the source, permitting to use real data with fake coordinates to accurately describe the background distribution of events.

For identifying the signal, discriminating variables are the direction of the reconstructed neutrino track and the energy proxy, $N_{\mbox{\tiny{HITS}}}$, whose normalised distributions are used as an input in a likelihood function as probability density functions (PDFs).
The signal PDF, $\mathcal{S}$, is built from simulated data weighted according to the WIMP annihilation spectra \cite{PPPC4}; the background PDF, $\mathcal{B}$, is obtained from shuffled data.
To assess the signal significance, a large number of skymaps (pseudo-experiments) are generated injecting an variable number of signal events, $n_s$, according to the signal PDF, over a set of $N = n_s+n_{bg}$ events, with $n_{bg}$ background events. The total number of events, $N$, is obtained from the total number of tracks in the data sample.
The	algorithm	used	to	search	for	an	excess	of	events	coming	from	the	region	of	the	Galactic	Centre	is	based	
on	an	unbinned	likelihood	function, $\mathcal{L}$, associated	with	each	skymap	
(containing $N$ events) 
\begin{equation}
\begin{split} 
    \log \mathcal{L} (n_s) = \sum_{i=1}^{N} \log\left[ n_s \mathcal{S}(\psi_i, N_{\mbox{\tiny{HITS}}}^i, q_i) \right. \\
    + \left. n_{bg} \mathcal{B}(\delta_i, N_{\mbox{\tiny{HITS}}}^i, q_i) \right]- n_{bg} - n_s,
\end{split}
\end{equation}
where $\psi_i$ is the angular distance of the $i$-th event from the Galactic Centre; $\delta_i$, the Equatorial declination of the $i$-th event; $N^i_{\mbox{\tiny{HITS}}}$, the number of light hits recorded by the detector and associated with the $i$-th reconstructed track, and $q_i$, the quality of the reconstruction.
The likelihood maximisation returns the number of signal events, $n_s^{*}$, found to belong to a cluster around the fixed coordinates of the Galactic Centre ($\alpha,\delta$) = (266$^\circ$,$-29.01^\circ$). 
The significance of a cluster is established by the test statistics, TS, which is a function of the ratio between the maximum and the pure background likelihood
\begin{equation}
    TS = -\log\frac{\mathcal{L} (n_s^{*})}{\mathcal{L}(n_s=0)}.
\end{equation}
To determine the significance of the observed TS, a series of pseudo-experiments is generated. This is performed by creating a large number of skymaps with a variable number of injected signal events, $n_s$, and running a maximum likelihood algorithm on each, returning the fitted number of events $n_s^*$ for each of them.
The number of events in each set of pseudo-experiments is subject to fluctuations following a Poisson distribution. To include this effect, a transformation through a Poisson function, $\mathcal{P}$, is performed, returning the TS as a function of the Poissonian mean $\mu$:
\begin{equation}
P\left(TS (\mu)\right) = \sum_{n^*_s=1}^N P\left(TS(n_s^*)\right)  \,\mathcal{P} (n_s^*, \mu),
\end{equation}
where $P(TS)$ indicates the TS distribution.

The main source of systematic uncertainties comes from the determination of the neutrino track direction. The track reconstruction relies on the time resolution of the detector, dependent on the photomultiplier time spread,
on the calibration and on possible space misalignment of the detector lines. The effect of systematic uncertainties was estimated in a previous analysis \cite{AntaresAF} to a total of 15\%.
A Gaussian smearing of  15\% is applied to the signal PDFs to account for detector systematics.


\subsection{Sensitivity of the search method}
\label{sectionsens}
Following Neyman's prescription~\cite{Neymanarticle}, an average upper limit on the number of signal events is computed from the median of the background test statistics $\overline{TS}_0$, 
compared with each distribution $P(TS)$ for each pseudo-experiment set. 
The sensitivity is defined as the 90$\%$ C.L. upper limit for a measurement equal to the median of the background TS distribution.
The analysis cuts are optimised to yield the best sensitivity (see Section \ref{sec2} and values reported in Table \ref{tablecut}).
If, after unblinding, a value smaller than the median of the background TS is observed in the data, limits are set equal to the sensitivity .

In case of a non-observation, a limit of the total number of signal events in the data ($\mu_{90}$) is converted into a limit on the integrated flux, $\Phi_{\nu+{\bar{\nu}}}$, through the acceptance, $\mathcal{A}$, and the livetime, $t$, as
\begin{equation}
\label{eq:flux}
     \Phi_{\nu+{\bar{\nu}}} = \frac{\mu_{90}}{\mathcal{A}\,\cdot t}.
\end{equation}
The acceptance is defined as the convolution of the effective area, $A_{eff}$~\cite{AntaresAF}, with each annihilation mode spectrum $dN_\nu/dE_\nu$~\cite{PPPC4}:
\begin{equation}
    \mathcal{A}(M) = \int_{E_{0}}^M A_{eff}^\nu (E_\nu) \frac{dN_\nu (E_\nu)}{dE_\nu} dE_\nu +
    [\nu\rightarrow\bar\nu],
\end{equation}
where $M$ is the considered WIMP mass, $E_{0}$ the energy threshold of the detector, determined from the first non-empty bin of the effective area, and 
$[\nu\rightarrow\bar\nu]$ indicates a symmetric term for antineutrinos.
The detector effective area increases with energy due to the raise with energy of the CC
cross section, combined with the better track definition 
of high-energy events, and with an increase in the muon range, making such that partially contained tracks can still be measured. 
The acceptance calculation relies on spectra provided by PPPC4.
The integrated flux of Equation~(\ref{eq:flux}) is converted into a measurement (limit) on the thermally averaged cross section for WIMP annihilation $\langle \sigma v \rangle$ using Equation~(\ref{flux}), for a given $J$-factor assuming a specific parameterisation of the DM halo model.

\begin{figure}
    \centering
    \includegraphics[width=0.7\textwidth]{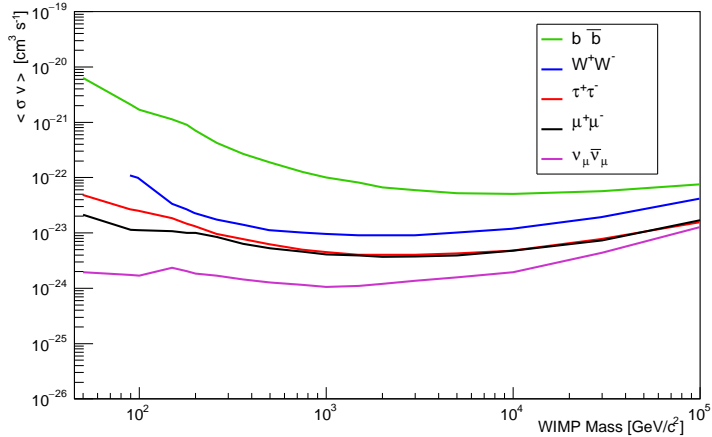}
\caption{Upper limits at 90$\%$ C.L. on the thermally averaged cross section for WIMP pair annihilation as a function of the WIMP candidate mass set with 11 years of ANTARES data, shown for five independent annihilation channels (each with 100\% branching ratio) and NFW halo model \cite{NFW}.}
    \label{fig:channels}
\end{figure}

\begin{figure}
    \centering
    \includegraphics[width=0.7\textwidth]{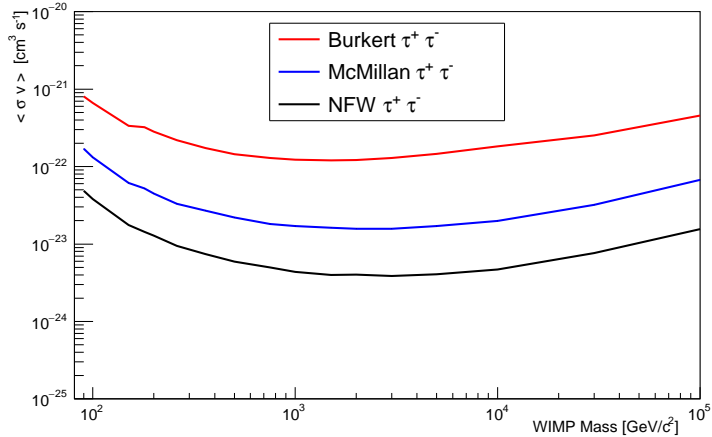}
    \caption{Upper limits at 90$\%$ C.L. on the thermally averaged cross section for WIMP pair annihilation as a function of the WIMP candidate mass set with 11 years of ANTARES data for three different halo models~\cite{NFW,McMillan,Burkert}. Here, only the $\tau^+\tau^-$ channel is shown.}
    \label{fig:halo}
\end{figure}

\section{Results}

Upon unblinding, the TS computed for 11 years of ANTARES data is compatible with background. We observed a TS smaller than the background median for all cases (masses and channels), hence we set all limit values equal to the corresponding sensitivities. This measurement sets limits on the cross section for WIMP-pair annihilation 
shown in Figure \ref{fig:channels} and computed according to Equation~(\ref{flux}). 
This figure shows limits for the five most prominent WIMP pair annihilation channels:
\begin{equation}
    \mbox{WIMP} \phantom{x} \mbox{WIMP}\rightarrow b \bar{b}, ~\tau^+ \tau^-, ~W^+ W^-, ~\mu^+ \mu^-, ~\nu{\bar{\nu}}
\end{equation}
independently computed with 100\% BR. 
The total amount of dark matter within a 30$^\circ$ angle around the Galactic Centre is taken into account, which corresponds to the solid angle $\Omega$ in Equation~(\ref{eqJ}). Best limits are obtained for the direct $\nu{\bar{\nu}}$ channel, as seen in Figure \ref{fig:channels}, which has the highest acceptance and the best sensitivity in number of events, due to the shape of the energy spectrum which peaks around the WIMP candidate mass; channels with steeply falling spectra such as $b\bar{b}$ give the least stringent limits.
Predictions on neutrino fluxes deriving from DM annihilation strongly rely on the parameterisation of the $J$-factor, as  mentioned in Section \ref{sec11}. Figure \ref{fig:halo} shows the $90\%$ C.L. limits on $\langle \sigma v \rangle$ for the $\tau^+\tau^-$ channel for three different halo models. The NFW profile~\cite{NFW} gives predictions over one order of magnitude more stringent than {\em{flat}} profiles such as Burkert \cite{Burkert}. An intermediate result is achieved 
for the McMillan profile \cite{McMillan} which has an intermediate inner slope. 
The results presented in this work represent an improvement ranging from a factor 1.1 to 1.4 with respect to the previous 9-year study \cite{ANTARES_GCWIMP}, according to the WIMP mass and channel considered.

\begin{figure}
    \centering
    \includegraphics[width=0.7\textwidth]{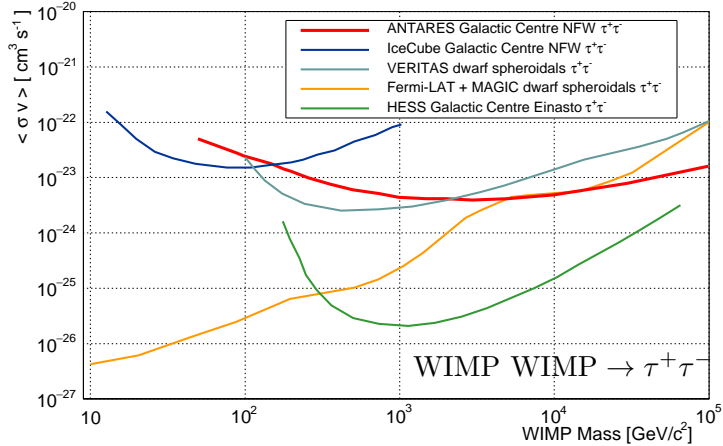}
\put(-140,27){$\mbox{WIMP} \mbox{ WIMP}\rightarrow \tau^+\tau^-$ }
    \caption{Limits on the thermally averaged cross section for WIMP pair-annihilation set with 11 years of ANTARES data, compared with current similar searches from IceCube \cite{IC86_GCWIMP} and from $\gamma$-ray telescopes HESS \cite{HESS}, VERITAS \cite{Veritas} and Fermi-LAT + MAGIC \cite{Fermi-MAGIC}. All curves are for the $\tau^+\tau^-$ benchmark channel.}
    \label{fig:others}
\end{figure}

\section{Discussion and conclusions}

Limits on the thermally averaged cross section $\langle \sigma v \rangle$ for DM annihilation towards the Galactic Centre were placed using 11 years of ANTARES data.
Some of the channels considered for this search also yield $\gamma \gamma$ pairs as a final product. For this case, ANTARES limits are set in context with existing limits from $\gamma$-ray telescopes (Figure \ref{fig:others}) for the $\tau^+\tau^-$ channel. In particular, the HESS Galactic Centre survey \cite{HESS} gives strong constraints thanks to the good visibility of this source from their location and to the prolongued observation campaign performed on this target. Note that both the MAGIC and the VERITAS detectors are located in the Northern Hemisphere and therefore they obtain their limits on the WIMP pair annihilation cross section from a campaign of observation of dwarf spheroidal Galaxies \cite{Veritas, Fermi-MAGIC}, not having the possibility to look directly into the Galactic Centre, if not with special settings for large zenith angle observations (e.g. ~\cite{MAGIClargeZenith}) with reduced sensitivity. 
Halo modeling in dwarf spheroidal Galaxies is subject to large uncertainties, and comparison with the Galactic Centre results is therefore not direct.
The results shown for IceCube \cite{IC86_GCWIMP} are obtained with Deep Core data, a configuration where the whole IceCube detector acts as a veto for atmospheric muons. Because of the Galactic Centre visibility, 
this analysis is limited to WIMP masses up to 1 TeV/c$^2$. 
All results shown in Figure \ref{fig:others} are obtained with the NFW profile, with the exception of the HESS result which refers to the Einasto DM halo model \cite{Einasto}.\\
The current searches for dark matter performed with ANTARES will be continued with
KM3NeT, which will instrument a total of about 1 km$^3$ of deep-sea water \cite{LOI}. KM3NeT has a modular layout consisting of blocks of 115 detection lines each. 
Two modules are being deployed in a large volume (36 m inter-optical-modules and 90 m inter-line spacing) to form the ARCA high-energy detector, and one in a denser  geometry instrumenting a smaller volume (9 m between optical modules and 20 m inter-line spacing) to form the ORCA low-energy detector.
As the prescriptions for the WIMP candidate mass vary over a broad range of values, both ARCA and ORCA will contribute to DM searches.

\section*{Acknowledgements}
The authors acknowledge the financial support of the funding agencies:
Centre National de la Recherche Scientifique (CNRS), Commissariat \`a
l'\'ener\-gie atomique et aux \'energies alternatives (CEA),
Commission Europ\'eenne (FEDER fund and Marie Curie Program),
Institut Universitaire de France (IUF), IdEx program and UnivEarthS
Labex program at Sorbonne Paris Cit\'e (ANR-10-LABX-0023 and
ANR-11-IDEX-0005-02), Labex OCEVU (ANR-11-LABX-0060) and the
A*MIDEX project (ANR-11-IDEX-0001-02),
R\'egion \^Ile-de-France (DIM-ACAV), R\'egion
Alsace (contrat CPER), R\'egion Provence-Alpes-C\^ote d'Azur,
D\'e\-par\-tement du Var and Ville de La
Seyne-sur-Mer, France;
Bundesministerium f\"ur Bildung und Forschung
(BMBF), Germany; 
Istituto Nazionale di Fisica Nucleare (INFN), Italy;
Nederlandse organisatie voor Wetenschappelijk Onderzoek (NWO), the Netherlands;
Council of the President of the Russian Federation for young
scientists and leading scientific schools supporting grants, Russia;
Executive Unit for Financing Higher Education, Research, Development and Innovation (UEFISCDI), Romania;
Ministerio de Ciencia, Innovaci\'{o}n, Investigaci\'{o}n y Universidades (MCIU): Programa Estatal de Generaci\'{o}n de Conocimiento (refs. PGC2018-096663-B-C41, -A-C42, -B-C43, -B-C44) (MCIU/FEDER), Severo Ochoa Centre of Excellence and MultiDark Consolider (MCIU), Junta de Andaluc\'{i}a (ref. SOMM17/6104/UGR), 
Generalitat Valenciana: Grisol\'{i}a (ref. GRISOLIA/2018/119), Spain; 
Ministry of Higher Education, Scientific Research and Professional Training, Morocco.
We also acknowledge the technical support of Ifremer, AIM and Foselev Marine
for the sea operation and the CC-IN2P3 for the computing facilities.

\bibliographystyle{unsrt}
\bibliography{ref}

\end{document}